\documentclass[conference]{IEEEtran}
\makeatletter
\let\@ORGmakecaption\@makecaption
\long\def\@makecaption#1#2{\@ORGmakecaption{#1}{#2}\vskip\belowcaptionskip\relax}
\makeatother

\usepackage[utf8]{inputenc}

\usepackage{cite}
\ifCLASSINFOpdf
  \usepackage[pdftex]{graphicx}
\else
\fi
\usepackage[cmex10]{amsmath}
\usepackage{amssymb}

\usepackage{array}
\ifCLASSOPTIONcompsoc
  \usepackage[caption=false,font=footnotesize,labelfont=sf,textfont=sf]{subfig}
\else
  \usepackage[caption=false,font=footnotesize]{subfig}
\fi
\usepackage{fixltx2e}
\usepackage{dblfloatfix}
\usepackage{lipsum}

\usepackage{url}
\usepackage[pdftex,unicode,hidelinks,bookmarks=false,draft]{hyperref}

\newcommand{\hreffootnote}[2]{\footnote{#2}}
 
\newcommand{\crefbrcmsmac}[2]{\hreffootnote{http://lxr.free-electrons.com/source/drivers/staging/brcm80211/brcmsmac/#1?v=2.6.39\#L#2}{brcmsmac source file: #1, kernel version: 2.6.39, line: #2}}
\newcommand{\crefutil}[2]{\hreffootnote{http://lxr.free-electrons.com/source/drivers/staging/brcm80211/util/#1?v=2.6.39\#L#2}{util source file: #1, kernel version: 2.6.39, line: #2}}

\usepackage{listings}
\usepackage{calc}
\newcommand*\phantomas[3][c]{% 
\ifmmode 
\makebox[\widthof{$#2$}][#1]{$#3$}% 
\else 
\makebox[\widthof{#2}][#1]{#3}% 
\fi 
}

\newif\ifblinded
\blindedfalse

\let\oldmarginpar\marginpar
\renewcommand\marginpar[1]{\-\oldmarginpar[\raggedleft\footnotesize #1]%
{\raggedright\footnotesize #1}}

\usepackage{ifthen}
\newboolean{draftversion}

\setboolean{draftversion}{false} %set false for removing comments and watermark

\ifthenelse{\boolean{draftversion}}{\setlength{\paperwidth}{10.5in}\setlength{\hoffset}{1in}}{}
\newcommand{\tdmh}[1]{\ifthenelse{\boolean{draftversion}}{{\color{orange}{\textbar}\hspace{-.4em}\marginpar{\color{orange}{\emph{MH:}\\#1}}}}{}}
\newcommand{\tdms}[1]{\ifthenelse{\boolean{draftversion}}{{\color{red}{\textbar}\hspace{-.4em}\marginpar{\color{red}{\emph{MS:}\\#1}}}}{}}
\newcommand{\tdsw}[1]{\ifthenelse{\boolean{draftversion}}{{\color{OliveGreen}{\textbar}\hspace{-.4em}\marginpar{\color{OliveGreen}{\emph{SW:}\\#1}}}}{}}
\newcommand{\tddsto}[1]{\ifthenelse{\boolean{draftversion}}{{\color{blue}{\textbar}\hspace{-.4em}\marginpar{\color{blue}{\emph{DSto:}\\#1}}}}{}}
\ifthenelse{\boolean{draftversion}}{\usepackage[firstpage]{draftwatermark}}{}

\newcommand\thefontsize[1]{{#1 The current font size is: \f@size pt\par}}

\usepackage{eso-pic}

\blindedfalse

\begin{document}

\title{NexMon: A Cookbook for Firmware Modifications on Smartphones to Enable Monitor Mode}

\author{\IEEEauthorblockN{Matthias~Schulz, Daniel~Wegemer and Matthias~Hollick}
\IEEEauthorblockA{Secure Mobile Networking Lab, TU Darmstadt, Germany}%
\IEEEauthorblockA{Email:\{\href{mailto:mschulz@seemoo.tu-darmstadt.de}{mschulz}, 
\href{mailto:dwegemer@seemoo.tu-darmstadt.de}{dwegemer}, 
\href{mailto:mhollick@seemoo.tu-darmstadt.de}{mhollick}\}@seemoo.tu-darmstadt.de}}%

\maketitle

% As a general rule, do not put math, special symbols or citations in the
% abstract
\begin{abstract}

Full control over a Wi-Fi chip for research purposes is often limited by its
firmware, which makes it hard to evolve communication protocols and test schemes
in practical environments. Monitor mode, which allows eavesdropping on all
frames on a wireless communication channel, is a first step to lower this
barrier. Use cases include, but are not limited to, network packet analyses,
security research and testing of new medium access control layer protocols.
Monitor mode is generally offered by SoftMAC drivers that implement the media
access control sublayer management entity (MLME) in the driver rather than in
the Wi-Fi chip. On smartphones, however, mostly FullMAC chips are used to reduce
power consumption, as MLME tasks do not need to wake up the main processor. Even
though, monitor mode is also possible in FullMAC scenarios, it is generally not
implemented in today's Wi-Fi firmwares used in smartphones.
This work focuses on bringing monitor mode to Nexus 5 smartphones to enhance the
interoperability between applications that require monitor mode and BCM4339
Wi-Fi chips. The implementation is based on our new C-based programming
framework to extend existing Wi-Fi firmwares.

\end{abstract}

\section{Introduction}

The use cases for open Wi-Fi firmwares or simply some firmware extensions like
monitor mode on a smartphone are numerous. Besides security research, it also
includes the implementation and testing of new medium access control (MAC) layer
protocols. Currently, either SoftMAC drivers are required, that give access to
the media access control sublayer management entity (MLME) or solutions such as
the Wireless MAC processor \cite{wirelessmacprocessor2012} that even allow to
modify the real time behaviour of a Wi-Fi chip. More flexibility is given by
connecting software-defined radios (SDRs) to a smartphones to even give access
to the 802.11 physical layer. Schulz et al. use this setup in \cite{Schulz2015}
to enhance wireless video transmissions by assigning physical modulation schemes
to different quality layers in a scalable video coding (SVC) scenario.
Nevertheless, SDR-based solutions are unlikely to be seen in mobile consumer
devices in the near future, as the energy consumption of field-programmable gate
arrays (FPGAs) used in SDRs is much higher than an energy-efficient single
purpose implementation.

Hence, in this work, we focus on extending the firmware of Wi-Fi chips used in
widely available consumer devices. Similar work to introduce monitor mode or
even frame injection has already been done for older chip generations in the
bcmon\cite{bcmon2013} (BCM4329 and BCM4330), as well as the monmob
\cite{monmob2012} (BCM4325 and BCM4329) projects. In this work, we focus on
BCM4339 chips used, for example, in the Nexus 5 smartphone. The contributions of
our work are as follows:

\begin{itemize}
  \item Reverse engineering of the BCM4339 chip and its firmware.
  \item Design and implementation of a C-based firmware extension framework, usable on firmwares for ARM processors.
  \item Firmware modification to support monitor mode on BCM4339 chips.
\end{itemize}

This work is structured as follows, in \autoref{sec:reverseengineering}, we
describe the reverse engineering process. In \autoref{sec:cframework}, we
describe our C-based framework for firmware modifications. In
\autoref{sec:framereception}, we describe how frame reception is handled in the
chip. In \autoref{sec:monitormode}, we document our monitor mode implementation.
In \autoref{sec:tryItYourself}, we explain how you use our project on your
phone. In \autoref{sec:knownBugsAndFutureWork}, we list bugs and future work and
conclude with \autoref{sec:conclusion}. The source code of our implementation
can be downloaded from \cite{nexmonproject2015}.

\begin{figure}[t!]
  \vspace{0.5cm}
  \centering
  \includegraphics[width=\linewidth]{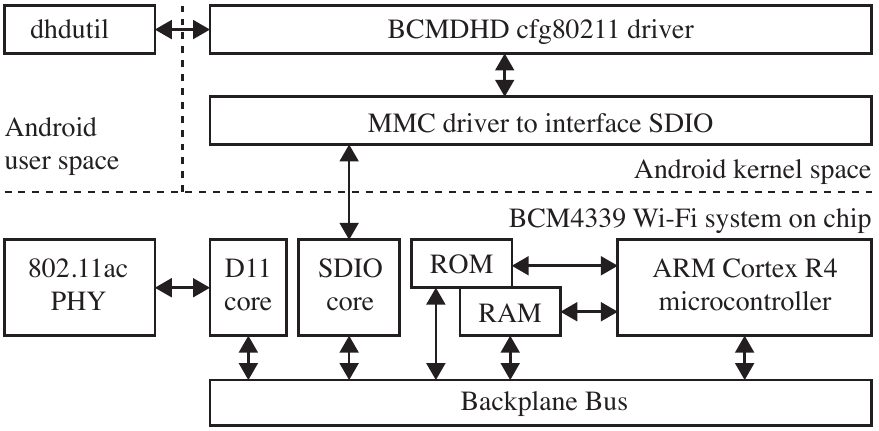}
  \caption{The BCM4339 system on chip is interfaced by the BCMDHD driver through SDIO. The driver itself can be controlled using the dhdutil.}
  \label{fig:systemonchip}
\end{figure}

\section{Reverse Engineering}
\label{sec:reverseengineering}

Before extending the existing Wi-Fi firmware, we had to analyze how the original
firmware works internally and how the system on chip in the Wi-Fi chip looks
like. We performed the analysis analogous to \cite{monmob2012} and
\cite{bcmon2013}. In \autoref{fig:systemonchip}, we illustrate the parts of the
system on chip relevant for this work. Our analysis is based on firmware version
6.37.32.RC23.34.40 (r581243) being delivered with Android 6.0 build MRA58K and
Android 6.0.1 build MMB29K for Nexus 5 smartphones.

\subsection{Explaining the system on chip}
\label{sec:explainingTheSystemOnChip}

On the Android side, the main component is the BCMDHD FullMAC driver that
interfaces the BCM4339 system on chip through the secure digital input output
(SDIO) interface. In general, it exchanges Ethernet frames with the Wi-Fi chip,
and the ARM microcontroller in the BCM4339 handles the re-framing into Wi-Fi
frames as well as the management of access point or device-to-device
connections. The real-time parts, like transmitting acknowledgement frames after
a fixed delay, are handled by the D11 core, that interfaces the 802.11ac
physical layer. Data between the different cores is exchanged over the backplane
bus. One can either write into registers or shared memory areas of the connected
cores or use direct memory access (DMA) controllers to exchange chunks of data
without involving the microcontroller.
The SDIO core has a DMA controller that is generally configured to exchange SDIO
frames between RAM and the SDIO controller. Whenever a new frame is copied to
RAM, the DMA controller triggers the \emph{external} interrupt at the
microcontroller, which then handles the received frame. The microcontroller can
also trigger the SDIO DMA to copy SDIO frames from RAM to the SDIO core for
transmission to the BCMDHD driver. The D11 core operates similarly: it has four
first in first out (FIFO) queues that can be accessed by DMA controllers to
exchange Wi-Fi frames. One FIFO is used for received and transmitted frames and
three additional FIFOs are used for the transmission of frames at different
quality levels (RX data and background data
packets\crefbrcmsmac{wlc\_bmac.c}{545}, best-effort data
packets\crefbrcmsmac{wlc\_bmac.c}{560}, video data
packets\crefbrcmsmac{wlc\_bmac.c}{574} and voice data and transmit-status
packets\crefbrcmsmac{wlc\_bmac.c}{585}).

\subsection{The D11 core}
\label{sec:theD11Core}

The D11 core consists of a programmable state machine (PSM) that is optimized
for real-time processing and can quickly change its program flow after checking
conditions in registers. This core is the first stage to decide, which frames
should be received or dropped. To allow monitor mode, the
maccontrol\crefbrcmsmac{d11.h}{134} register needs to be set to activate
promiscuous mode\crefbrcmsmac{d11.h}{466}. The activation can be performed by
the microcontroller by calling the
wlc\_bmac\_mctrl\crefbrcmsmac{wlc\_bmac.c}{1483} function. How, we identified
such a function in the existing firmware is described in the following.

\subsection{The microcontroller firmware analysis}

The microcontroller is used to act as an interface between the different cores
in the system on chip and the BCMDHD driver in the Linux kernel. It implements a
SoftMAC driver similar to brcmsmac which is part of Broadcom's
brcm80211\hreffootnote{http://lxr.free-electrons.com/source/drivers/staging/brcm80211/?v=2.6.39}{http://lxr.free-electrons.com/source/drivers/staging/brcm80211/?v=2.6.39}
driver. Our code references in the footnotes, hence, often refer to this driver
source code. Besides the SoftMAC implementation, the microcontroller's firmware
also performs re-framing, as well as additional tasks to
unburden the main smartphone processor and enhance energy efficiency. To
implement monitor mode, we intend to bypass these additional features, but first
we need to start analysing the firmware itself.

As illustrated in \autoref{fig:systemonchip} and described in \cite{bcmon2013}
and \cite{monmob2012}, the firmware consists of a part that resides in ROM
(size: 640\,KB, starting at 0x0) and another one that is loaded by the BCMDHD
driver into RAM (size: 768\,KB, starting at 0x180000). The firmware file for the
RAM is found as a binary file on the smartphone. The firmware that resides in
ROM can simply be extracted using the
dhdutil\hreffootnote{https://android.googlesource.com/platform/hardware/broadcom/wlan/+/master/bcmdhd/dhdutil/}{https://android.googlesource.com/platform/hardware/broadcom/wlan/+/\\master/bcmdhd/dhdutil/}
that is distributed as part of the Android platform and allows to send specific
ioctls to the BCMDHD driver in case the driver was compiled with active
DHD\_DEBUG flag. Using the membytes command, one can read at arbitrary memory
locations in ROM and RAM and dump the output to a binary file.

To analyse the extracted firmware, we load it into the Interactive Disassembler
(IDA) and use Hex-Rays Decompiler for ARM to be able to convert assembler code
into more readable C-like code which we can compare to the brcmsmac's source
code to find similarities. To identify functions, we first go through the list
of detected strings and can identify function names from the brcmsmac driver.
Cross referencing code that uses these strings leads to constructions that look
like printf function calls resulting in an output in the console that can be
read using dhdutil's consoledump command. The printf calls look as follows:
\begin{lstlisting}
printf("%s: some error...", __FUNCTION__);
\end{lstlisting}%
First the name of the surrounding function is printed then an error message.
Analysing all function name strings, we are able to name many functions. To name
even more functions, we look for functions in the brcmsmac source code that call
already named functions or that are called by already named functions. To better
understand the code, it is very helpful to classify variable types and create
structs in IDA. This made us discover the \lstinline|wlc->pub->_cnt| struct that
contains counters for various statistics, which help up to classify the
surrounding code. Before continuing to describe how we found the path from an
interrupt triggering a new frame reception and the transmission to the BCMDHD
driver, we introduce our framework that allows to write firmware patches in C instead of
Assembler.

\section{C-based Firmware Modification Framework}
\label{sec:cframework}

After creating the first patches for the firmware to output debug information
and to get a first monitor mode prototype running in Assembler, we decided that
programming in C is less error prone, hence, we developed a C-based programming
framework. We intended to generate functions that can replace already existing
functions in the firmware or that can be called before calling an existing
function by redirecting branch instructions to our own functions. Additionally,
we wanted to directly call existing functions in the firmware.

\setcounter{lstlisting}{1} 
\begin{lstlisting}[float=*b, caption=Disassembly of the dma\_rx hooking hello world program. The binary was created using our C-based programming framework., label=lst:helloworldasm]
ROM:00180000                 PUSH            {R4,LR}
ROM:00180002                 MOVS            R4, R0
ROM:00180004                 LDR             R0, =aHelloWorld ; "hello world"
ROM:00180006                 BL              0x126F0 ; printf
ROM:0018000A                 MOVS            R0, R4
ROM:0018000C                 BL              0x8C69C ; dma_rx
ROM:00180010                 POP             {R4,PC}
ROM:00180010 ; ---------------------------------------------------------------------------
ROM:00180012                 ALIGN 4
ROM:00180014 off_180014      DCD aHelloWorld         ; DATA XREF: ROM:00180004r
ROM:00180014                                         ; "hello world"
ROM:00180018 aHelloWorld     DCB "hello world",0     ; DATA XREF: ROM:00180004o
ROM:00180018                                         ; ROM:off_180014o
\end{lstlisting}

In our implementation, we define external function prototypes of firmware
functions we like to call and use a linker to insert correct branch instructions
and place global variables in memory. Unfortunately, the simple definition of
symbols with addresses of functions in memory leads to the creation of
trampoline stubs to switch from Thumb to ARM instruction set, even though the
existing firmware code needs to be executed in Thumb mode. To work around this
problem, we created an object file with dummy functions for all firmware
functions, we wanted to call. Each of those functions is placed in a separate
section and---using the linker---can be placed at the appropriate position where
the code resides in the firmware.

Using this approach, the generated machine code is position dependent and uses
branch with link instructions relative to the current program counter. This
allows small binaries and calls to functions like printf, which can take an
arbitrary number of parameters.

The hello world program in \autoref{lst:helloworld} serves as an example. It is
implemented as a hook to the dma\_rx\crefutil{hnddma.c}{732} function and prints
``hello world'' to the console, which can be read using dhdutil's consoledump
command. The resulting assembler code after disassembling in IDA is illustrated
in \autoref{lst:helloworldasm}. Using our framework, we can rewrite any firmware
function we like, which we do to activate monitor more. Before that, we first
need to understand which functions are called when receiving a frame.

\section{Frame reception}
\label{sec:framereception}

In this section, we describe how a Wi-Fi frame reception is handled by the
microcontroller. As mentioned above, the DMA controller of the D11 core
transfers received frames into RAM and then triggers the \emph{external}
interrupt of the microcontroller. The microcontroller handles this interrupt by
settings its program counter to execute the fast interrupt (FIQ) instruction in
the exception vector that is located at address 0x0 in the ROM. In the BCM4339
ROM this instruction is a branch to an exception handler located in RAM at
0x180fee, which branches to a common exception handler at 0x181032, which calls
the callback function referenced at 0x181100, which points to 0x181e48. In case
of a fast interrupt it calls the function at 0x181a88, which calls each function
of a linked list of functions. The reference to the list is placed at 0x180e5c.
The first function pointer in this list is a path to
wlc\_dpc\crefbrcmsmac{wlc\_bmac.c}{313} (0x27550 $\Rightarrow$ 0x2733c
$\Rightarrow$ 0x61eb4), which checks if new frame receptions from the D11 core
need to be handled using the wlc\_bmac\_recv\crefbrcmsmac{wlc\_bmac.c}{257}
function (0x1aad98), which is called through its wrapper at 0x4f7a4.

The wlc\_bmac\_recv function is mentioned in the bcmon project as a good
starting point to implement monitor mode. It consists of three parts. At the
beginning it calls the dma\_rx\crefutil{hnddma.c}{732} function in a loop to
collect pointers to the received frames in a list of linked sk\_buffs. Then it
calls dma\_rxfill\crefutil{hnddma.c}{809} to allocate new receive buffers and
post them to the DMA ring buffer so that they can be filled with new frames. In
the last part, wlc\_bmac\_recv handles the frames stored in the previously
created linked list. The main frame processing starts in the
wlc\_recv\crefbrcmsmac{wlc\_bmac.c}{6993} function at 0x19afe8. The processing
result is a call to dngl\_sendpkt (named according to the bcmon project
presentation) following for example the path 0x19955f $\Rightarrow$ 0x198cdd
$\Rightarrow$ 0x1981f5 $\Rightarrow$ 0x1893b5 $\Rightarrow$ 0x183771
$\Rightarrow$ 0x182C84 $\Rightarrow$ 0x182750. This function takes an sk\_buff
and an SDIO channel number, prepends the buffer's data payload with an SDIO
frame header and passes the buffer on to a dma\_txfast\crefutil{hnddma.c}{1426}
call (0x18256c $\Rightarrow$ 0x182450 $\Rightarrow$ 0x1844b2) that uses the
SDIO's DMA to transfer the frame to the BCMDHD driver. How this processing pass
can be bended to bypass the frame processing and directly pass raw frames to the
BCMDHD driver is explained in the next section.

\setcounter{lstlisting}{0} 
\begin{lstlisting}[float=t, caption=Hello world program hooking to the dma\_rx function., label=lst:helloworld]
#include "wrapper.h"

struct sk_buff *dma_rx_hook(struct dma_info *di) {
	printf("hello world");
	return dma_rx(di);
}
\end{lstlisting}

\section{Monitor Mode}
\label{sec:monitormode}

Our approach to enable monitor mode is twofold. First, we replace the
wlc\_bmac\_recv function to bypass frame processing and directly forward the
received raw frames to the BCMDHD driver. Second, we correctly set the
maccontrol register during the initialization of the interface to work in
promiscuous mode.

\subsection{Rewriting the wlc\_bmac\_recv function}

Our rewritten wlc\_bmac\_recv function from \autoref{lst:wlcbmacrecv} can be
compiled with our C-based framework to get a replacement for the existing
wlc\_bmac\_recv function in the firmware. Compared to the wlc\_bmac\_recv
function of the brcmsmac source code, our function has an additional parameter
cnt that saves the number of loop iterations with calls to dma\_rx. Instead of
first calling dma\_rx in a loop to build a linked list of sk\_buffs, we decided
to fetch frames from the D11 DMA and directly transmit them over the SDIO
interface using the dngl\_sendpkt function. The constant SDIO\_INFO\_ADDR is the
address of a struct that holds information on the SDIO core. To avoid stalling
the interrupt handling for too long, if many Wi-Fi frames are received, we limit
the number of processed frames to bound\_limit and return whether we stayed below
that limit.

\subsection{Setting the correct maccontrol registers}

As mentioned above, the maccontrol registers in the D11 core need to be set
appropriately to receive all frames. The monmob developers already identified
the wlc\_bmac\_mctrl\crefbrcmsmac{wlc\_bmac.c}{1483} (0x4f080) function to
perform this task. To activate the reception of all frames at the initialization
of the wireless interface, we set the registers in the
wlc\_coreinit\crefbrcmsmac{wlc\_bmac.c}{2300} (0x1ab66c) function, which calls
wlc\_bmac\_mctrl with the parameters mask at 0x1ab82c and value at 0x1ab828.
We extend mask and value with the appropriate bits that are also set by bcmon.
Somehow, the original firmware resets those bits during operation, hence, we set
them again during each execution of the wlc\_bmac\_recv function (see
\autoref{lst:wlcbmacrecv}).

\section{Try it yourself}
\label{sec:tryItYourself}

\begin{figure}[t!]
  \centering
  \includegraphics[width=\linewidth]{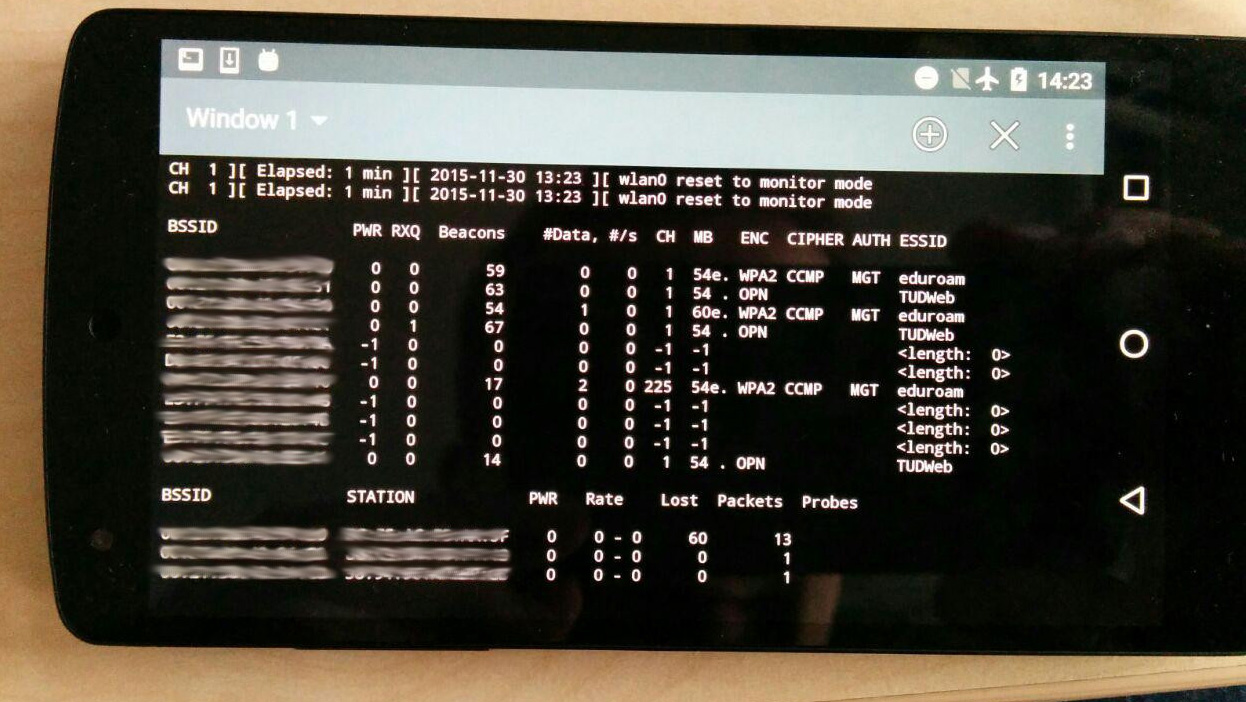}
  \caption{Running airodump-ng on the Nexus 5 in monitor mode.}
  \label{fig:airodump}
\end{figure}

To let you try our monitor mode implementation on your Nexus 5 smartphone, we
provide a boot.img that contains a custom kernel based on the
android-msm-hammerhead-3.4-marshmallow-mr1\hreffootnote{https://android.googlesource.com/kernel/msm/+/android-msm-hammerhead-3.4-marshmallow-mr1}{https://android.googlesource.com/kernel/msm/+/android-msm-hammerhead-3.4-marshmallow-mr1}
branch with modules enabled and the BCMDHD driver compiled as a module. Our
driver modification is based on the bcmon project. The module is located in
/nexmon/bcmdhd.ko and can be loaded with insmod and unloaded with rmmod. It is
not automatically loaded during startup. Additionally, we disabled the
wpa\_supplicant and p2p\_supplicant services in the init.hammerhead.rc file,
hence, regular Wi-Fi does not work with our provied boot image. The monitor
interface is broad up by executing \lstinline|ifconfig wlan0 up|.
The received frames contain a radiotap header that can be processed by programs
that rely on monitor mode enabled interfaces. To test the monitor capabilities
and change channels, we included pre-compiled binaries of iwconfig and tcpdump
under /nexmon/bin/. The result of running another tool called airodump-ng on the
phone is shown in \autoref{fig:airodump}.

\emph{Be advised that using our boot.img may damage your phone and may void your
phone's warranty! You use our tools at your own risk and responsibility!} To run
the boot.img on your own Nexus 5, you should have Android 6.0.1 build MMB29K
installed and an unlocked bootloader. \emph{Be aware that unlocking the
bootloader generally wipes your phone!} To play with monitor mode we advise you
to boot our image instead of flashing it, as it disables regular Wi-Fi
capabilities. Simply execute 
\lstinline|adb reboot bootloader && fastboot boot boot.img| 
or run \lstinline|make boot|, in case you cloned our git repository.
If you intend to modify the firmware on your own, you can compile everything
from source using the Makefile in our git repository \cite{nexmonproject2015}.

\begin{lstlisting}[float=t, caption=Rewritten wlc\_bmac\_recv function to directly send raw Wi-Fi frames to the BCMDHD driver., label=lst:wlcbmacrecv]
#include "bcm4339.h"
#include "wrapper.h"

int wlc_bmac_recv(struct wlc_hw_info *wlc_hw, 
	unsigned int fifo, int bound, int *cnt) {
	void *p;
	int n = 0;
	int bound_limit = wlc_hw->wlc->pub->tunables->rxbnd;

	do {
		if((p = dma_rx(wlc_hw->di[fifo]))) {
			dngl_sendpkt(SDIO_INFO_ADDR, p, 0xF);
		}
		n++;
	} while(n < bound_limit);
	*cnt += n;
	
	dma_rxfill(wlc_hw->di[fifo]);
	
	wlc_bmac_mctrl(wlc_hw, (MCTL_PROMISC | MCTL_KEEPCONTROL | MCTL_BCNS_PROMISC | MCTL_KEEPBADFCS), (MCTL_PROMISC | MCTL_KEEPCONTROL | MCTL_BCNS_PROMISC | MCTL_KEEPBADFCS));
	
	if(n < bound_limit) {
		return 0;
	} else {
		return 1;
	}
}
\end{lstlisting}

\section{Known Bugs and Future Work}
\label{sec:knownBugsAndFutureWork}

Currently not all received frames contain valid Wi-Fi frames and the radiotap
headers are set to default values. Additionally, the Wi-Fi firmware crashes
under unknown circumstances, which requires a chip reset by running
\lstinline|ifconfig wlan0 down && ifconfig wlan0 up|. In future releases we
intend to fix those bugs and implement frame injection to be able to transmit
arbitrary Wi-Fi frames.

\section{Conclusion}
\label{sec:conclusion}

This work describes how to analyse and successfully extend Wi-Fi firmwares to
enable monitor mode on a BCM4339 FullMAC Wi-Fi chip. To simplify firmware
modifications, we present a C-based firmware extension framework and use it in
our implementation. The functionality of our approach is demonstrated by running
airodump-ng on a Nexus 5 smartphone.

% \newpage use section* for acknowledgement
\section*{Acknowledgment} \ifblinded Acknowledgment has been blinded for
review.\hfill\\\vspace*{4em} \else This work has been funded by the German
Research Foundation (DFG) in the Collaborative Research Center (SFB) 1053 ``MAKI
– Multi-Mechanism-Adaptation for the Future Internet'' and by LOEWE CASED, which
provided the IDA Pro and HexRays Decompiler licenses. Many thanks go to Andrés
Blanco who supported us in our reverse engineering phase.
% Many thanks go to our shepherd, Srdjan Capkun.
\fi

\IEEEtriggeratref{8}

\vspace{13pt}
%\clearpage
%\nocite{*}
%\newpage
\bibliographystyle{IEEEtranS}
\bibliography{IEEEabrv,bibliography}

%\noindent linewidth: \the\linewidth\\
%textwidth: \the\textwidth

% \clearpage

% \section*{Font Sizes}
% \thefontsize\tiny
% \thefontsize\scriptsize
% \thefontsize\footnotesize
% \thefontsize\small
% \thefontsize\normalsize
% \thefontsize\large
% \thefontsize\Large
% \thefontsize\LARGE
% \thefontsize\huge

% \section*{Widths}
% text width: \the\textwidth \par
% line width: \the\linewidth \par

\end{document}